\documentclass[sigconf]{acmart}
\usepackage{booktabs}
\usepackage{multirow}
\usepackage{graphicx}
\usepackage[normalem]{ulem}
\AtBeginDocument{%
  }

\copyrightyear{2026}
\acmYear{2026}
\setcopyright{cc}
\setcctype{by}
\acmConference[CIKM '26]
  {Proceedings of the 35th ACM International Conference on Information and Knowledge Management}
  {November 07--11, 2026}
  {Rome, Italy}
\acmBooktitle{Proceedings of the 35th ACM International Conference on Information and Knowledge Management (CIKM '26), November 07--11, 2026, Rome, Italy}
\acmDOI{10.1145/3799682.3841065}
\acmISBN{979-8-4007-2539-5/2026/11}
\definecolor{darkgreen}{rgb}{0.0, 0.45, 0.0}

\begin{document}

\title{Closing the Long-Short View Gap in Sequential Recommendation without Cached History}

\author{Lingfeng Shi}
\authornote{Corresponding authors.}
\affiliation{%
  \institution{Texas A\&M University}
  \city{College Station}
  \country{United States}}
\email{lingfengs111@tamu.edu}

\author{Chengkai Huang}
\authornotemark[1]
\affiliation{%
  \institution{University of New South Wales}
  \city{Sydney}
  \country{Australia}}
\email{chengkay.huang@gmail.com}

\author{Lina Yao}
\affiliation{%
  \institution{University of New South Wales}
  \city{Sydney}
  \country{Australia}}
\email{lina.yao@unsw.edu.au}

\author{James Caverlee}
\affiliation{%
  \institution{Texas A\&M University}
  \city{College Station}
  \country{United States}}
\email{caverlee@tamu.edu}


\renewcommand{\shortauthors}{Lingfeng Shi, Chengkai Huang, Lina Yao, and James Caverlee}


\begin{abstract}
Sequential recommenders are typically trained on long user histories to capture rich behavioral signals, yet serving with training-length sequences is often impractical due to real-time efficiency constraints. Directly using only recent behaviors leads to a severe performance drop. To bridge this gap, existing approaches compress user histories into persistent per-user states, storing and retrieving them at inference time; while effective, they impose non-trivial infrastructure overhead and offer little remedy in cold-start scenarios. In this paper, we empirically identify two structural flaws rooted in geometric properties and dataset sparsity, and propose a novel two-stage framework to close the long-short-view performance gap. Specifically, in the first stage, we replace the commonly used dot-product with angular similarity scoring and leverage a modified softmax to counter prefix position bias. In the second stage, we fine-tune only bias and LayerNorm components—universal to standard sequential backbones—for further improvement. Both stages are guided by carefully designed learning objectives. Extensive experiments on two representative backbones across three public datasets demonstrate the effectiveness of our proposed framework.
\end{abstract}




\maketitle

\section{Introduction}
\begin{figure}[t]
    \centering
    \includegraphics[width=\linewidth]{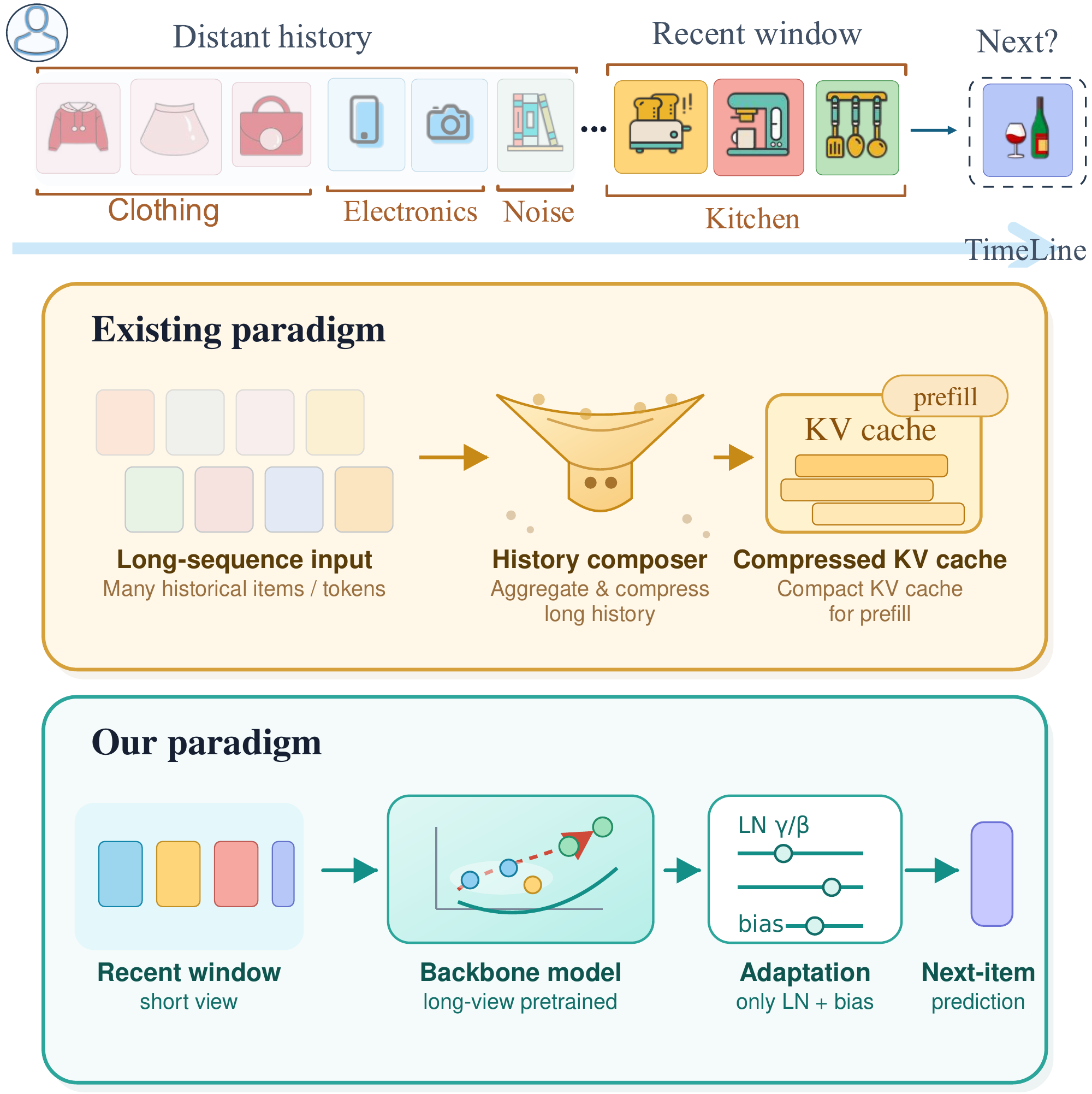}
    \caption{Comparison of two paradigms for sequential recommendation inference. Existing methods compress long user histories into KV cache or token memory for prefill; our method instead pretrains the backbone on long-view data and serves with only a short recent window, adapting LayerNorm and bias parameters solely.}
    \Description{A two-part pipeline comparison. The existing paradigm compresses a long sequence into a persistent KV cache used for prefill. The proposed paradigm feeds only a recent short window to a long-view-pretrained backbone and adapts only LayerNorm and bias parameters before next-item prediction.}
    \label{fig:intro_tradeoff}
\end{figure}
Sequential recommendation predicts a user's next interaction from an ordered history of past behaviors~\cite{pan2026survey}. 
Among existing paradigms, ID-based sequential recommendation remains foundational to large-scale web systems, where users and items are represented by learned embeddings indexed by discrete identifiers. 
Because these models are structurally lightweight, fast at inference, and straightforward to deploy, optimizing their capacity to handle expanding contexts has attracted substantial research interest~\cite{liu2024mamba4rec,zhai2024actions,wang2025rethinking}. 

Richer historical contexts provide more behavioral evidence, and recent long-sequence models have demonstrated that extending the history window can significantly improve recommendation quality~\cite{zhu2024lifelong,chai2025longer,liao2025multi}. However, this long-history advantage hardly carries over to serving time. In production, due to real-time latency requirements, most systems can only operate on a user's relatively short, recent interactions~\cite{si2024twin,chang2023twin,agrawal2023sarathi}. When the model is consequently restricted to this narrow window at inference, recommendation quality degrades sharply~\cite{chang2023twin,lv2019sdm}, exposing a profound \textit{long-short-view performance gap}.

A natural response to this gap is to bring the distant past into the current request: representative approaches compress user history into learnable tokens, KV caches, or recurrent states that are precomputed offline and retrieved at serving time~\cite{zhang2025efficient,chen2026recurrent,chu2026kwai}. These methods transfer long-range signals effectively, but they require per-user historical states to be materialized, stored, and kept current, trading one form of infrastructure cost for another~\cite{zhang2025efficient,wang2026mtserve}.
Moreover, these methods still depend on per-user state being available at serving time, leaving the cases where per-user state is unavailable---such as new users or users returning after a long absence---entirely unaddressed~\cite{pi2019practice,luo2026awakening}.
Motivated by these facts, we ask a different question: \textit{Rather than carrying the long history forward to the serving system, can we train a backbone that remains effective when inference is restricted to the short recent window alone?}

Through empirical analysis, we identify two structural flaws that compound the long-to-short degradation (Figure~\ref{fig:intro_tradeoff}).
First, dot-product scoring entangles directional alignment with embedding magnitude---a \textit{norm bias} that tracks item popularity in sparse long-context datasets (Sec.~\ref{sec:amplified_sparse}) and becomes a miscalibrated shortcut under input truncation. Second, we observe that softmax attention tends to concentrate residual weight on early prefix positions, whose abrupt removal at short-view inference disrupts the learned attention pattern---an empirical regularity analogous to the sinking phenomenon reported in deeper models~\cite{xiao2023efficient}.

Guided by these findings, we propose a two-stage backbone-agnostic framework. As is shown in the bottom part of Fig~\ref{fig:intro_tradeoff}, the first stage targets the long-view pretraining and counters geometric flaws: we replace dot-product scoring with angular similarity and adopt bounded attention aggregation ($\operatorname{softmax}_1$)~\cite{miller2023attention}.
The second stage closes the remaining length-induced distribution gap: we fine-tune only the pre-existing bias and LayerNorm interfaces of the backbone---without introducing any new parameters---under a joint objective of pointwise alignment and cross-view distillation from the long-view model.

To sum up, our contributions are as follows:
\begin{itemize}
    \item We study a practical sequential recommendation setting where 
    models are trained on long user histories but serve with only 
    a short recent window. Empirically, we identify two structural flaws driving the long-short-view gap: dot-product norm bias---a popularity shortcut amplified by dataset sparsity---and prefix attention bias, whose disruption under input truncation causes performance degradation.
    \item We propose a backbone-agnostic two-stage framework for short-view inference without cached history. In the long-view training stage, angular similarity scoring and bounded attention aggregation ($\operatorname{softmax}_1$) 
    correct the geometric flaws. In the short-view adaptation stage, bias/LayerNorm-only fine-tuning under a joint objective closes the remaining distribution gap---requiring no cached tokens.
    \item Experiments on SASRec and HSTU across three public datasets demonstrate that our framework surpasses full-sequence baselines under short-view-only inference; with bias/LayerNorm adaptation, it remains competitive with cache- and token-based methods--- without any per-user precomputation or storage. Codes are released\footnote {\url{https://github.com/lingfengs111/long-short-view-rec}}.
\end{itemize}

\section{Preliminaries}

\subsection{Sequential Recommendation Setup}
Sequential recommendation aims to give a ranked list based on the candidate item and the user's historical interactions, e.g., like clicks, purchases, or views. Formally, let $\mathcal{I}$ denote the item set. Each user is associated with a chronologically ordered interaction sequence:
\begin{equation*}
\mathbf{X_u} = [x_1, x_2, \ldots, x_t], \qquad x_t \in \mathcal{I}.
\label{eq:prelim-seq}
\end{equation*}
Here, $x_1$ represents the earliest item the user interacted with, while $x_t$ represents the most recent item. Given the observed sequence $\mathbf{X_u} = [x_1, \ldots, x_t]$, the task is to predict the next item $x_{t+1}$. 

In our settings, an ID-based sequential backbone $f_{\theta}$ encodes the user history sequence into a $d$-dimensional user state:
\begin{equation}
\mathbf{h}_t = f_{\theta}(\mathbf{x}_{u}) \in \mathbb{R}^{d},
\label{eq:prelim-backbone}
\end{equation}
and each candidate item $i \in \mathcal{I}$ is associated with an embedding $\mathbf{e}_i \in \mathbb{R}^{d}$. The notation $\mathbf{h}_t$ can be regarded as the query representation of a specific user.

In ID-based recommender systems, the common paradigm in the next step is to score every candidate item $i$ with the state of the user query using the raw dot product~\cite{rendle2020neural,steck2024cosine}:
\begin{equation}
z_t(i) = \mathbf{h}_t^{\top}\mathbf{e}_i.
\label{eq:prelim-dot}
\end{equation}
After that, $z_t(i)$ is assigned to a final value via a specific loss function, and finally the ranked list can be generated by simply sorting these results.

\subsection{Long- and Short-View Setup}
The distinction between long and short sequences forms a core concept of this paper: given a user sequence $\mathbf{X_u} = [x_1, x_2, \ldots, x_t]$, we chronologically partition it into three segments: the items closest to the user's next interaction are termed \textbf{\textit{the recent}} segment; the oldest interactions constitute \textbf{\textit{the prefix}} segment; and the remaining intermediate interactions form \textbf{\textit{the middle}} segment. Typically, both \emph{the prefix} and \emph{the recent} segments consist of only a few items, whereas \emph{the middle} segment spans a significantly longer history. This partition reflects a functional distinction rather than an arbitrary length boundary: each segment plays a qualitatively different role in the backbone's attention mechanism, as our empirical analysis in Sec.~\ref{sec:analysis} will reveal.

Let $L$ denote the maximum window size supported by a given system. We define the \textbf{\textit{Long-View}} history as:
\begin{equation}
\mathbf{x}^{(L)}
=
[x_1, \ldots, x_L]
=
\mathbf{x}_{prefix} \parallel \mathbf{x}_{middle} \parallel \mathbf{x}_{recent},
\label{eq:prelim-view-defs}
\end{equation}
and the \textbf{\textit{Short-View}} history as:
\begin{equation}
\mathbf{x}^{(S)}
=
[x_{L-r+1}, \ldots, x_L]
=
\mathbf{x}_{recent},
\label{eq:prelim-view-short}
\end{equation}
which keeps only the recent segment used during fine-tuning and inference.

We therefore focus on \textbf{training} on \textbf{\textit{Long-View}} data, followed by \textbf{fine-tuning and inference} on \textbf{\textit{Short-View}} data.

\subsection{Dataset Setup} \label{dataset_setup}
To guide the following empirical analysis, we next introduce the dataset setup. Specifically, we adopt two widely recognized industrial e-commerce datasets, 
Taobao\footnote{\url{https://tianchi.aliyun.com/dataset/649}} and XLong\footnote{\url{https://tianchi.aliyun.com/dataset/22482}}, and one popular public movie dataset MovieLens-10M\footnote{\url{https://grouplens.org/datasets/movielens/10m/}} as our base datasets. 

We follow the leave-one-out (LOO) protocol: for each user sequence, the last interaction is held out for testing, the second-to-last for validation, and all preceding interactions form the training set. Since most user sequences in the original datasets are relatively short and fail to meet the requirements of a long-context setting, we apply a data processing strategy following~\cite{feng2024long,chen2026recurrent}. Specifically, for MovieLens and Taobao, we retain user sequences with a training length greater than 200 (202 total); for XLong, we keep sequences longer than 400 (402 total). Detailed dataset statistics can be found in Table~\ref{tab:dataset_statistics}.

\section{Empirical Analysis}
\label{sec:analysis}

In production recommender systems, encoding hundreds of historical interactions per request is computationally expensive and latency-sensitive, making short-view inference a practical necessity. Yet a model trained to converge on long-view data is implicitly optimized for that richer input, so naively switching to a short-view context at inference time is expected to degrade recommendation quality. This tension motivates our core research question: \textit{\uline{Given a high-quality model trained to convergence on long-view data, how can we maintain comparable recommendation performance at inference time using exclusively short-view data as input}}?

To address this question, we start with two sub-questions. \textbf{RQ1}: For a well-fitted long-view model, to what extent does the recommendation performance drop if we directly feed short-view data during inference? \textbf{RQ2}: What are the distinct impacts on recommendation quality when input segments at different chronological positions are restored? Addressing these core questions guides the unfolding of our subsequent analysis.

\subsection{Zero-Shot Inference and Recovery} \label{zs_infer}
\begin{figure}[t]
\centering
\includegraphics[width=\linewidth]{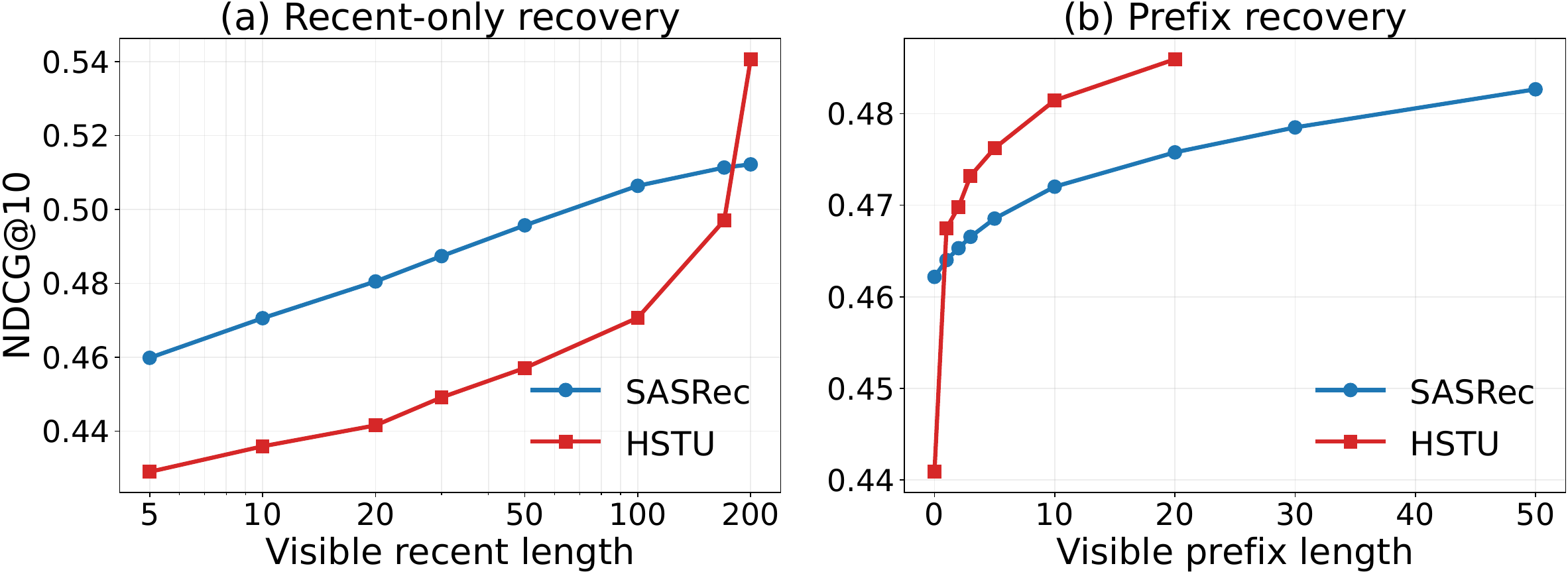}
\caption{Trained on Taobao dataset. \textbf{(a)} Recent-only recovery under different suffix lengths. \textbf{(b)} Prefix recovery: keep the short-view length fixed and restore different prefix tokens.}
\Description{Two line charts for SASRec and HSTU on Taobao. Panel (a) shows NDCG at 10 increasing as the visible recent suffix grows toward the full 200-item history. Panel (b) shows a sharp initial recovery when a few prefix tokens are restored, followed by smaller continued gains.}
\label{fig:motivation-main}
\end{figure}

In order to answer \textbf{RQ1} and \textbf{RQ2}, we select two representative ID-based recommendation models: SASRec~\cite{kang2018self}, a widely adopted causal self-attention model for sequential recommendation, and HSTU~\cite{zhai2024actions}, a recent industrial-scale backbone featuring a gated pointwise-aggregated attention mechanism designed for large-scale deployment. We train both models on long-view data $\mathbf{x}^{(L)}$ until convergence with their own best training recipe. Then we freeze them and directly test them with short-view data $\mathbf{x}^{(S)}$. 

Figure~\ref{fig:motivation-main}(a) plots NDCG@10 as the visible recent length grows from 5 to 200 tokens. At the shortest window of 5 tokens, both models already achieve non-trivial performance (SASRec $\approx 0.48$, HSTU $\approx 0.44$), confirming that recommendations are primarily driven by the most recent interactions. As the window expands, performance improves monotonically for both backbones, but the recovery curves differ in shape: SASRec recovers smoothly and near-linearly, while HSTU remains relatively flat until around 50 tokens and then rises sharply, suggesting its attention mechanism benefits more from a sufficiently long recent context. Both models only approach their full-view performance ceiling when the window reaches the training length $L=200$, indicating that a persistent gap remains for any short-view window—the gap that our adaptation method aims to close.

Figure~\ref{fig:motivation-main} (b) discloses another recovery pattern: a drastic performance improvement is observed when directly feeding the model with the concatenation of \textbf{\textit{the prefix}} segment and \textbf{\textit{the recent}} segment, i.e. $\mathbf{x}_{prefix} \parallel \mathbf{x}_{recent}$, while preserving each token's own positional encoding. Furthermore, once the number of prefix tokens exceeds 5, it turns into less pronounced but continuous growth. Similar patterns have been discussed in~\cite{xiao2023efficient, chai2025longer}. The sharp initial jump in panel~(b) is a manifestation of the \textbf{\textit{sinking effect}}: because softmax forces attention weights to sum to one, models learn to route residual attention onto early prefix tokens as structural ``dump'' positions rather than for their content. Removing these tokens at short-view inference disrupts the learned pattern, causing an immediate performance drop even when the missing prefix carries little semantic value.

Another distinct observation is that, excluding the sharp leap from sinking tokens, the remaining tokens across both segments yield a steady, linear performance recovery. This suggests that tokens at other positions do not store unique information by virtue of their placement. 


These findings directly answer \textbf{RQ1} and \textbf{RQ2}: short-view inference incurs a meaningful but recoverable performance gap, and the three segments play qualitatively distinct roles in closing it. However, if we aim to bridge this gap and recover long-view effects using exclusively short-view data, adapting a tailored parameter subset is a necessary first step. This motivates our third research question \textbf{RQ3}: How can we formulate an inherent compatibility between the representation geometry and the interaction mechanism to optimally calibrate short-view adaptation?

\subsection{Amplified Sparsity and Norm Bias} \label{sec:amplified_sparse}
\begin{table}[t]
\centering
\small
\setlength{\tabcolsep}{6pt}
\renewcommand{\arraystretch}{1.2}
\begin{tabular}{lccc}
\hline
Dataset & Freq.\ $=1$ items & Freq.\ $\le 5$ items & $\rho(\mathrm{freq}, \|\mathbf{e}_i\|_2)$ \\
\hline
ML-10M & $1.16\%$ & $11.60\%$ & $-0.93$ \\
Taobao & $37.98\%$ & $72.98\%$ & $0.54$ \\
XLong & $51.69\%$ & $84.86\%$ & $0.80$ \\
\hline
\end{tabular}
\vspace{6pt}
\caption{Item-frequency concentration and its relation to item embedding norm in a standard dot-product SASRec.}
\label{tab:norm-frequency-main}
\vspace{-10pt}
\end{table}
To address \textbf{RQ3}, we examine the characteristics of the datasets directly. Following the preprocessing protocols in Sec.~\ref{dataset_setup}, we report dataset statistics in Table~\ref{tab:dataset_statistics} and low-frequency item statistics in Table~\ref{tab:norm-frequency-main}. The latter reveals that all three processed datasets exhibit prominent sparsity—an issue that is particularly severe in the industrial-dataset Taobao and XLong, whereas ML-10M remains comparatively dense.

Since an item's frequency attribute is often heavily linked to its geometric properties, we compute the \textbf{\textit{Pearson correlation coefficient}} between item frequencies and their corresponding embedding norms derived from the converged backbone model. The empirical results, summarized in Table~\ref{tab:norm-frequency-main}, demonstrate that a robust correlation between these two variables is clearly evident across all scenarios, despite the fact that embedding norms do not always strictly increase with frequency (as exemplified by the inverse decreasing trend on ML-10M).

This matters: when ${\sim}80\%$ of items are poorly modeled, the large norms of a few high-frequency items act as a popularity shortcut in dot-product scoring. Expanding Eq.~\eqref{eq:prelim-dot}:
\begin{equation}
z_t(i) = \mathbf{h}_t^{\top}\mathbf{e}_i = \|\mathbf{h}_t\|_2 \, \|\mathbf{e}_i\|_2 \, \cos(\theta_{t,i}),
\label{eq:analysis-dot-decomp}
\end{equation}
reveals that $\|\mathbf{h}_t\|_2$ and $\|\mathbf{e}_i\|_2$ can dominate the ranking score, overriding the directional alignment $\cos(\theta_{t,i})$.

While popularity signals utilized by dot-product routing are generally effective~\cite{rendle2020neural,steck2024cosine}, they introduce shortcuts and training instability in long-context recommendation settings, where dataset sparsity and uneven item support are largely amplified. Therefore, to learn a cleaner embedding geometry tailored for long-view to short-view representation transfer, prioritizing directional alignment becomes a more well-suited paradigm for this sparse, long-context scenario. This establishes the geometric principle behind \textbf{RQ3}: in sparse long-context settings, ranking should be governed by directional alignment rather than magnitude.
 
\section{Proposed Method}
\label{sec:method}

\noindent
The framework unfolds in two phases. The first corrects the geometric flaws identified in Sec.~\ref{sec:analysis}---norm bias and prefix sinking---through changes to backbone structure and scoring. The second closes the remaining length-induced distribution gap via parameter-efficient short-view adaptation.

\subsection{Long-View Backbone Learning}
\label{sec:backbone-learning}

The backbone model is trained on full long-view sequences $\mathbf{x}^{(L)}$ to establish the foundational representation space for item ranking. 

To achieve this, our backbone design incorporates three synergistic components: (a) a magnitude-agnostic directional scorer to eliminate frequency bias; (b) a modified attention mechanism that normalizes prefix positions to moderate the "sinking effect”; and (c) a standard sampled softmax loss, empirically selected as the optimal objective function for this stage.

\subsubsection{Angular Similarity Scoring}
Inspired by the empirical insights in Sec.~\ref{sec:amplified_sparse}, the final ranking rule should be governed primarily by directional alignment rather than vector magnitude. Crucially, this normalized embedding geometry, rooted in the embedding table, serves as the structural knowledge that the short-view student inherits and aims to preserve. To achieve this, we substitute the raw dot product with a temperature-scaled cosine similarity:
\begin{equation}
\begin{aligned}
s_t(i)
&=
\frac{\bar{\mathbf{h}}_t^{\top}\bar{\mathbf{e}}_i}{\tau}, \\
\bar{\mathbf{h}}_t
&=
\frac{\mathbf{h}_t}{\|\mathbf{h}_t\|_2},
\qquad
\bar{\mathbf{e}}_i
=
\frac{\mathbf{e}_i}{\|\mathbf{e}_i\|_2},
\end{aligned}
\label{eq:method-sim}
\end{equation}
where $s_t(i)$ denotes the similarity score assigned to candidate item $i$ at position $t$, and $\tau > 0$ denotes the temperature hyperparameter. By normalizing both the history representation and item embeddings, we entirely decouple vector magnitude from the ranking score. Consequently, the model is compelled to express user preferences exclusively through angular alignment, effectively preventing the representation space from degenerating into a popularity-driven shortcut.

While cosine scoring has been explored to alleviate popularity bias~\cite{zhai2023revisiting}, we specifically tailor it for the long-context setting, where dataset sparsity and skewed item support are amplified by prolonged sequence lengths. As Sec.~\ref{exp_overall} shows, this magnitude-agnostic scoring consistently retains long-view backbone effectiveness under short-view inference.


\subsubsection{Bounded Attention Aggregation}
To counteract the prefix-induced sinking effect highlighted in Sec.~\ref{zs_infer}, we target its architectural origin. Instead of adding shared global tokens~\cite{feng2024long} or user profiles augmenting the prefix, we resolve this issue by modifying the attention aggregation mechanism itself.

Specifically, let $a_{t,j}$ denote the attention logit from position $t$ to token $j \le t$. Standard softmax forces these logits to sum to one across all visible positions. We instead consider $\operatorname{softmax}_1$ ~\cite{miller2023attention}, which introduces a constant term in the denominator:
\begin{equation}
\begin{aligned}
\alpha_{t,j}^{\operatorname{softmax}}
&=
\frac{\exp(a_{t,j})}{\sum_{m \le t} \exp(a_{t,m})}, \\[4pt]
\alpha_{t,j}^{\operatorname{softmax}_{1}}
&=
\frac{\exp(a_{t,j})}{1 + \sum_{m \le t} \exp(a_{t,m})}.
\end{aligned}
\label{eq:method-softmax1}
\end{equation}
The rationale is that it empowers the model to allocate near-zero total attention mass when no visible tokens are particularly informative, rather than forcing it to distribute a fixed unit probability budget. This flexibility is important in our scenario; as demonstrated by our prior analysis, the sparsity inherent in long-context recommendation implies that the next-item prediction could rely on merely a few specific tokens, even the latest token itself. 

The effectiveness of this design is validated by our subsequent experiments. Notably, although the original HSTU ~\cite{zhai2024actions} advocates for a new pointwise aggregated attention variant, our empirical results reveal that $\text{softmax}_1$ exhibits a superior capacity for short-view adaptation. Therefore, we standardize on $\text{softmax}_1$ across architectures.

\subsubsection{Backbone Optimization Objective} \label{Backbone_Optimi}
We train the backbone with one positive item $x_{t+1}$ and a sampled negative set $\mathcal{N}_t$ using sampled softmax, which has been validated as an effective function~
\cite{klenitskiy2023turning}:
\begin{equation}
\mathcal{L}_{\mathrm{SS}}
=
-\log
\frac{
\exp\!\left(s_t(x_{t+1})\right)
}{
\exp\!\left(s_t(x_{t+1})\right)
+
\sum_{i \in \mathcal{N}_t}
\exp\!\left(s_t(i)\right)
}.
\label{eq:method-ss}
\end{equation}
Sampled softmax places the positive and negatives in direct competition within a shared denominator, which is a natural fit for the retrieval objective: the model should rank the target above alternatives in the same context, rather than solving a collection of independent binary decisions.

\subsection{Short-View Adaptation}

\subsubsection{Parameter-Type Selection: Bias and LayerNorm Only}
When reviewing the core question in Sec.~\ref{sec:analysis}, from long-view training to short-view inference, the change in sequence lengths could cause a radical shift in LayerNorm statistics. This phenomenon shares a conceptual parallel with our analysis in Sec.~\ref{sec:amplified_sparse}; where both issues center on how uncalibrated normalization boundaries-- whether at an item's vector-norm level or at sequence-wise level -- distort the representation geometry. Drawing inspiration from BitFit~\cite{zaken2022bitfit}, we focus our adaptation exclusively on the backbone's native LayerNorm layers and bias components.

Let $\theta$ denote all the parameters from the backbone model, then it could be decomposed to:
\begin{equation}
\theta = \theta_{\mathrm{frozen}} \cup \theta_{\mathrm{adapt}},
\qquad
\theta_{\mathrm{adapt}}
\subseteq
\mathcal{B} \cup \mathcal{A}_{\mathrm{LN}},
\qquad
\mathcal{A}_{\mathrm{LN}}=\{\gamma_{\mathrm{LN}}, \beta_{\mathrm{LN}}\}.
\label{eq:method-bitfit}
\end{equation}
where $\mathcal{B}$ and $\mathcal{A}_{\mathrm{LN}}$ represent the native bias tensors and LayerNorm affine parameters. This selection provides both multiplicative scaling and additive recalibration to absorb the length-induced distribution shift, serving as a pure parameter-type filter without introducing any auxiliary modules.
\begin{figure}[t]
\centering
\includegraphics[width=\linewidth]{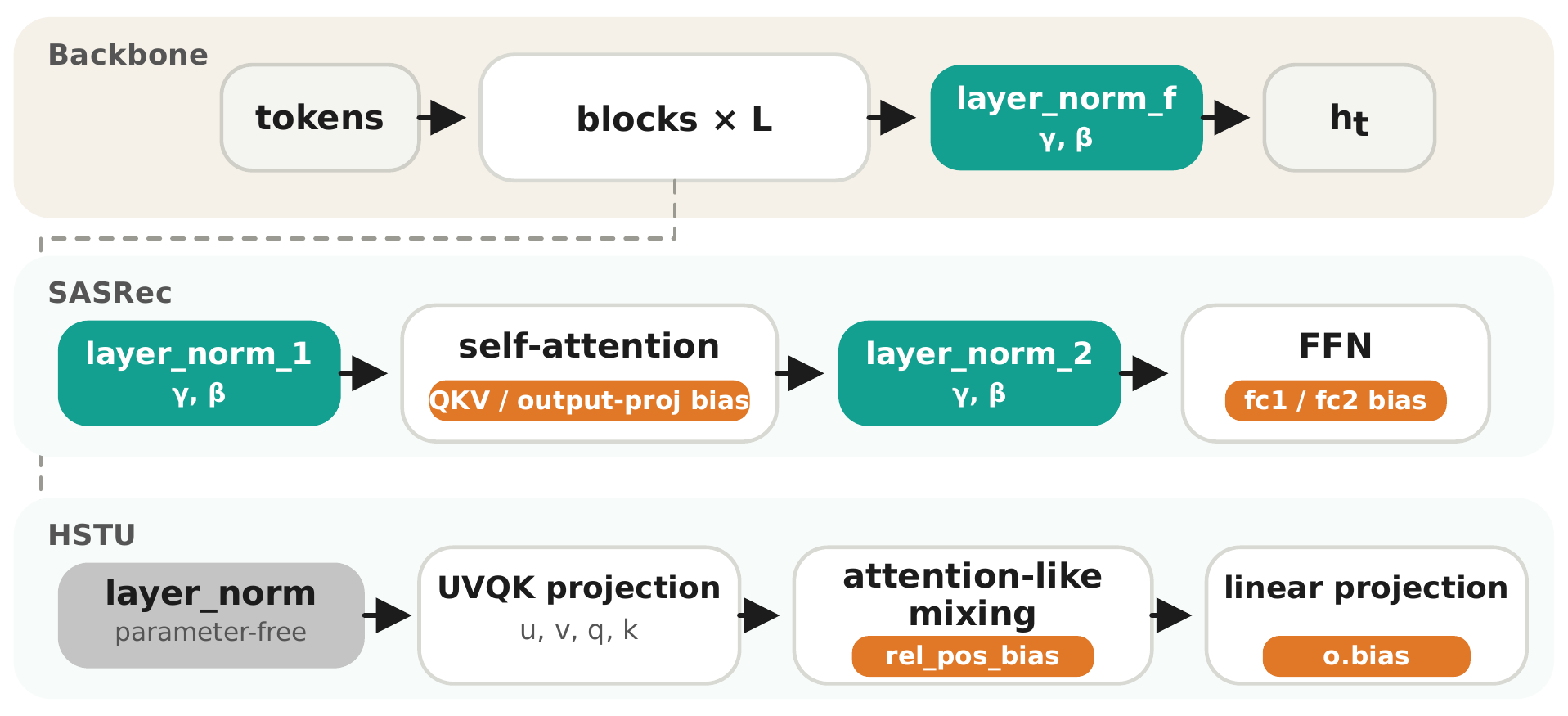}
\caption{Illustration of architecture-aware adaptation. Green squares and orange blocks represent LayerNorm modules and native bias components, respectively.}
\Description{A schematic of a sequential backbone highlighting the small parameter subset updated during adaptation. LayerNorm modules are marked in green and native bias components in orange, while all remaining backbone parameters stay frozen.}
\label{fig:adaptation-surface-map}
\end{figure}

As illustrated in Figure~\ref{fig:adaptation-surface-map}, this rule automatically induces customized adaptation footprints based on the backbone's architecture. It heavily targets the distributed normalization scales ($\gamma, \beta$) in LayerNorm-reliant models, while seamlessly shifting its focus to projection and positional biases in more lightweight, parameter-free normalization models. 

Crucially, this design represents a precise and localized adaptation strategy. Improper choices could trigger representation collapse, yielding short-view data finetuning results inferior to a minimal baseline. By freezing the robust backbone weights and adjusting only the bias-norm interfaces, we preserve the foundational long-view knowledge while accommodating the short-view input dynamics.

\subsubsection{Short-View Adaptation Objective}
\label{sec:shortview-obj}

Given short-view input $\mathbf{x}^{(S)}_t$, ranking scores are computed identically to Eq.~\eqref{eq:method-sim}:
\begin{equation*}
\mathbf{h}^{(S)}_t = f_{\theta}\!\left(\mathbf{x}^{(S)}_t\right),
\qquad
s^{(S)}_t(i) = \bar{\mathbf{h}}^{(S)\top}_t\bar{\mathbf{e}}_i\,/\,\tau.
\end{equation*}
We supervise the adapted parameters with a binary cross-entropy (BCE) loss over one positive--negative pair, a foundational objective in classic sequential models~\cite{kang2018self}:
\begin{equation}
\mathcal{L}_{\mathrm{BCE}}
=
-\Bigl[
\log \sigma\!\left(s^{(S)}_t(x_{t+1})\right)
+\log\!\left(1 - \sigma\!\left(s^{(S)}_t(i)\right)\right)
\Bigr],
\label{eq:method-ft-bce}
\end{equation}
where $i$ is a uniformly sampled negative item. Notably, we empirically find that re-applying sampled softmax during adaptation yields negligible performance gains; the absolute pointwise alignment of BCE offers a more direct and fine-grained objective, making it highly compatible with this stage.

\subsubsection{Cross-View Augmentation}
\label{sec:kd}

To further enrich adaptation with global context, the corresponding long-view sequence can be naturally introduced as a complementary behavioral view. Specifically, the long-view sequence $\mathbf{x}^{(L)}_t$ is processed to project a ranking score:
\begin{equation*}
\mathbf{h}^{(L)}_t = f_{\theta}\!\left(\mathbf{x}^{(L)}_t\right),
\qquad
s^{(L)}_t(i) = \bar{\mathbf{h}}^{(L)\top}_t\bar{\mathbf{e}}_i\,/\,\tau.
\end{equation*}



To effectively transfer the structural relative ordering and the full predictive probability distribution inherent in the long-view context, we project both views onto a shared localized candidate pool comprising the ground-truth item $x_{t+1}$ and a set of randomly sampled negatives $\mathcal{N}_t$. We then minimize the Kullback-Leibler (KL) divergence between their respective softmax distributions over this joint pool:
\begin{equation}
\begin{aligned}
\mathcal{L}_{\mathrm{KD}}
= \operatorname{KL}\!\Bigl(
&\operatorname{softmax}\!\Bigl(\bigl[s^{(L)}_t(x_{t+1}),\,\{s^{(L)}_t(i)\}_{i\in\mathcal{N}_t}\bigr] / T_d\Bigr),\\
&\operatorname{softmax}\!\Bigl(\bigl[s^{(S)}_t(x_{t+1}),\,\{s^{(S)}_t(i)\}_{i\in\mathcal{N}_t}\bigr] / T_d\Bigr)
\Bigr),
\end{aligned}
\label{eq:method-kd}
\end{equation}
where $T_d > 0$ is the distillation temperature. 

Conceptually, this approach aligns with self-distillation paradigms~\cite{zhao2026self,chen2026onesearch}, where a richer model view guides a more constrained one.

\subsubsection{Dual-Supervision Objective}
\label{sec:dual-obj}

The two supervision signals are combined into a single fine-tuning objective:
\begin{equation}
\mathcal{L}_{\mathrm{FT}}
=
\mathcal{L}_{\mathrm{BCE}}
+
\lambda_{\mathrm{KD}} \mathcal{L}_{\mathrm{KD}},
\label{eq:method-ft-final}
\end{equation}
where $\lambda_{\mathrm{KD}} \ge 0$ balances direct target supervision against teacher-side candidate structure.

\section{Experiment}

We organize experiments around four questions: \textbf{(Q1)}~Backbone recipe gains; \textbf{(Q2)}~Adaptation effectiveness and parameter choice; \textbf{(Q3)}~Efficiency versus compression alternatives; and \textbf{(Q4)}~Temporal robustness of adaptation gains.

\subsection{Experimental Setup}
We first introduce the general experimental setup and key terminology to support a clearer interpretation of the results that follow.

\begin{table}[t]
\centering
\small
\caption{Dataset statistics after selection}
\setlength{\tabcolsep}{5.5pt}
\renewcommand{\arraystretch}{1.2}
\begin{tabular}{lccccc}
\hline
Dataset & Users & Items & Context len & Avg freq & Density \\
\hline
ML-10M & 10,870 & 9,841 & 202 & 223.1 & $2.05{\times}10^{-2}$ \\
Taobao & 94,489 & 2.06M & 202 & 9.25 & $9.79{\times}10^{-5}$ \\
XLong & 20,000 & 1.81M & 402 & 4.45 & $2.23{\times}10^{-4}$ \\
\hline
\end{tabular}
\vspace{6pt}

\label{tab:dataset_statistics}
\vspace{-10pt}
\end{table}

\subsubsection{Backbone Models}
We instantiate the same training and adaptation framework on two representative backbone models:
\begin{itemize}
\item \textbf{SASRec}~\cite{kang2018self}. We follow the recommended setting, instantiating the model with 2 attention blocks, one single attention head, and a hidden dimension of 128, setting the dropout rate to 0.1.
\item \textbf{HSTU}~\cite{zhai2024actions}. We follow the recommended setting, setting a 4-block architecture with 4 attention heads, where the hidden size remains at 128, and the dimensions per-head \(U/V\) and \(Q/K\) are set to \(32\), with a dropout rate of 0.2.
\end{itemize}

\subsubsection{Backbone-Training Settings}
As described in Sec.~\ref{sec:backbone-learning},the backbone model is trained on full long-view sequences $\mathbf{x}^{(L)}$  to
establish the foundational representation space. Specifically, we compare two backbone-training regimes:
\begin{itemize}
\item \textbf{Original.} The backbone's native training recipe, thereby aligning with the objective used in its original formulation. For example, SASRec utilizes binary cross-entropy (BCE) loss based on dot-product scoring, whereas HSTU employs sampled-softmax training.
\item \textbf{Ours.} Following the improvements proposed in Sec.~\ref{sec:backbone-learning}, our experiments incorporate angular similarity scoring with bounded attention.
\end{itemize}

\subsubsection{Evaluation Scenarios}
Here, we delineate several experimental scenarios to evaluate the impact of different context configurations.
\begin{itemize}
\item \textbf{Full.} The model is trained and evaluated on full-length sequences $\mathbf{x}^{(L)}$ ($L=200$ for ML-10M and Taobao, $L=400$ for XLong). This establishes the empirical performance upper bound. It is deployable but expensive, as encoding hundreds of historical items per request incurs substantially higher inference latency and compute cost.
\item \textbf{Short.} The model is trained on full-length sequences but evaluated on truncated short sequences (fixed length 20) during inference, without any adaptation. This is the zero-shot transfer baseline and represents a deployable setting, though it may incur a performance gap relative to \textbf{Full}.
\item \textbf{Adapted.} The same as \textbf{Short} at inference time, but with an additional lightweight fine-tuning stage on short-view data before deployment. This is our primary practical target: it retains the serving efficiency of \textbf{Short} while closing the performance gap through targeted parameter adaptation.
\item \textbf{+prefix.} A short-view inference scenario augmented with a small historical prefix. The recent context is set to 15 items and the prefix to 5 items, keeping the total input length at 20 for a fair comparison. This scenario tests how much value early-history signals add when they are available, at the cost of retrieving and storing a small number of historical interactions.
\end{itemize}

    

\subsubsection{Existing Methods}
In relevant sequence compression tasks, there are many related models or methods. Here, we select two representative methods and adapt them to our experimental settings for a fair comparison.
\begin{itemize}
\item \textbf{LONGER}$^{\dagger}$~\cite{chai2025longer}. An architecture that augments a standard sequence encoder with global anchor tokens and a grouped token-merge layer for long-history compression. We adapt its components to our ID-only LOO setting: item-ID sequences only, trained from scratch without warm-starting from our backbones, with 4 global tokens, merge size 4, and one inner layer. Comparisons are therefore limited to the SASRec backbone.\footnote{LONGER's InnerTrans assumes a standard transformer block (Q/K/V projections + FFN). HSTU instead uses a joint UVQK projection that couples gating and attention dimensions in a single operation, which is central to its efficiency design. Directly grafting InnerTrans onto HSTU would require restructuring this core block, breaking the efficiency properties that motivate HSTU's formulation.}
\item \textbf{PersRec}~\cite{zhang2025efficient}. This method represents approaches that learn gist tokens or summary tokens within input tokens. Here, we employ checkpoints from the backbone trained on long-view data as a warm start, and then fine-tune it with short sequences. Note that the scope of fine-tuning here covers \textbf{all parameters and learnable tokens}. In our main runs, PersRec uses $8$ learnable tokens, the recent segment length is also $20$ items, and an NCE-style objective. During inference, a mask is used to achieve a caching effect.
\end{itemize}

\subsubsection{Evaluation Protocol}
We use Hit Ratio (Hit@5, Hit@10) and NDCG (NDCG@5, NDCG@10) as the evaluation metrics. All results are evaluated under the same sampled-ranking protocol with $1000$ uniformly sampled negatives per target item. All experiments were conducted on a single server equipped with four NVIDIA RTX A5000 GPUs.

\subsubsection{Implementation Details}
All models are trained with AdamW. For our long-view backbone recipe, we use sampled-softmax supervision with $128$ sampled negatives, $\ell_2$-normalized user/item embeddings, temperature $0.07$, a peak learning rate of $10^{-3}$, a minimum learning rate of $10^{-5}$, and zero weight decay. SASRec uses cosine decay, while HSTU uses cosine decay with a $100$-step warmup. We train each backbone for up to $200$ epochs and select checkpoints by validation performance.

For the second-stage short-view adaptation, we initialize from the corresponding full-view checkpoint and optimize only the architecture-selected bias and LayerNorm parameters with AdamW at a learning rate $10^{-4}$ and zero weight decay. We train SASRec adaptations for up to $200$ epochs and HSTU adaptations for up to $120$ epochs, using validation performance for model selection.

\subsection{Overall Performance} \label{exp_overall}

\begin{table*}[t]
\centering
\caption{Recommendation performance across three datasets and two backbone architectures. \textbf{Ours} applies our long-view training recipe (angular scoring + bounded attention); \textbf{Original} uses each backbone's native recipe. $\Delta$ reports the absolute NDCG@10 change relative to \textbf{Original Full} of the same dataset. Bold marks the best result per column. $^\dagger$~LONGER is adapted to our ID-only LOO setting (item-ID sequences only, no user profiles or cross features). ``--'' indicates the configuration does not apply to this backbone.}
\label{tab:main-results}

\renewcommand{\arraystretch}{1.28}
\setlength{\tabcolsep}{3.6pt}
\newcommand{\mainmetric}[1]{\makebox[4em][c]{#1}}

\resizebox{\textwidth}{!}{%
\begin{tabular}{ccc|ccccc|ccccc}
\hline
\multirow{2}{*}{Dataset}
& \multirow{2}{*}{Group}
& \multirow{2}{*}{Setting}
& \multicolumn{5}{c|}{SASRec}
& \multicolumn{5}{c}{HSTU} \\
\cline{4-13}
& &
& \mainmetric{Hit@5} & \mainmetric{Hit@10} & \mainmetric{NDCG@5} & \mainmetric{NDCG@10} & $\Delta$
& \mainmetric{Hit@5} & \mainmetric{Hit@10} & \mainmetric{NDCG@5} & \mainmetric{NDCG@10} & $\Delta$ \\
\hline

\multirow{8}{*}{ML-10M}
& \multirow{3}{*}{Original}
& Full
& 0.2866 & 0.4037 & 0.1952 & 0.2329 & {\scriptsize\textit{ref}}
& 0.3169 & 0.4373 & 0.2172 & 0.2559 & {\scriptsize\textit{ref}} \\
& & Short
& 0.2677 & 0.3793 & 0.1797 & 0.2157 & {\scriptsize\textcolor{red}{$-1.7\%$}}
& 0.2717 & 0.3911 & 0.1832 & 0.2217 & {\scriptsize\textcolor{red}{$-3.4\%$}} \\
& & +prefix
& 0.2764 & 0.3939 & 0.1875 & 0.2255 & {\scriptsize\textcolor{red}{$-0.7\%$}}
& 0.2970 & 0.4201 & 0.2036 & 0.2432 & {\scriptsize\textcolor{red}{$-1.3\%$}} \\
\cline{2-13}

& \multirow{3}{*}{Ours}
& Full
& \textbf{0.3284} & \textbf{0.4470} & \textbf{0.2292} & \textbf{0.2696} & {\scriptsize\textcolor{darkgreen}{$\mathbf{+3.7\%}$}}
& \textbf{0.3793} & \textbf{0.4977} & \textbf{0.2698} & \textbf{0.3083} & {\scriptsize\textcolor{darkgreen}{$\mathbf{+5.2\%}$}} \\
& & Short
& 0.3116 & 0.4270 & 0.2163 & 0.2535 & {\scriptsize\textcolor{darkgreen}{$\mathbf{+2.1\%}$}}
& 0.3153 & 0.4320 & 0.2204 & 0.2580 & {\scriptsize\textcolor{darkgreen}{$\mathbf{+0.2\%}$}} \\
& & Adapted
& 0.3218 & 0.4392 & 0.2242 & 0.2633 & {\scriptsize\textcolor{darkgreen}{$\mathbf{+3.0\%}$}}
& 0.3393 & 0.4567 & 0.2387 & 0.2768 & {\scriptsize\textcolor{darkgreen}{$\mathbf{+2.1\%}$}} \\
\cline{2-13}

& \multirow{2}{*}{Others}
& LONGER$^{\dagger}$
& 0.2642 & 0.3675 & 0.1730 & 0.2093 & {\scriptsize\textcolor{red}{$-2.1\%$}}
& -- & -- & -- & -- & -- \\
& & PersRec
& 0.2881 & 0.4028 & 0.1904 & 0.2313 & {\scriptsize\textcolor{darkgreen}{$+0.1\%$}}
& 0.3268 & 0.4528 & 0.2227 & 0.2674 & {\scriptsize\textcolor{darkgreen}{$+1.3\%$}} \\
\hline

\multirow{8}{*}{Taobao}
& \multirow{3}{*}{Original}
& Full
& 0.5820 & 0.6479 & 0.4957 & 0.5175 & {\scriptsize\textit{ref}}
& 0.5878 & 0.6516 & 0.5020 & 0.5231 & {\scriptsize\textit{ref}} \\
& & Short
& 0.5439 & 0.6087 & 0.4621 & 0.4831 & {\scriptsize\textcolor{red}{$-3.4\%$}}
& 0.5211 & 0.5902 & 0.4222 & 0.4446 & {\scriptsize\textcolor{red}{$-7.9\%$}} \\
& & +prefix
& 0.5628 & 0.6304 & 0.4775 & 0.4995 & {\scriptsize\textcolor{red}{$-1.8\%$}}
& 0.5526 & 0.6182 & 0.4540 & 0.4741 & {\scriptsize\textcolor{red}{$-4.9\%$}} \\
\cline{2-13}

& \multirow{3}{*}{Ours}
& Full
& 0.6144 & 0.6677 & \textbf{0.5329} & 0.5500 & {\scriptsize\textcolor{darkgreen}{$+3.3\%$}}
& 0.6418 & 0.6992 & \textbf{0.5576} & \textbf{0.5764} & {\scriptsize\textcolor{darkgreen}{$\mathbf{+5.3\%}$}} \\
& & Short
& 0.6002 & 0.6522 & 0.5241 & 0.5407 & {\scriptsize\textcolor{darkgreen}{$\mathbf{+2.3\%}$}}
& 0.6031 & 0.6631 & 0.5201 & 0.5396 & {\scriptsize\textcolor{darkgreen}{$\mathbf{+1.7\%}$}} \\
& & Adapted
& 0.6190 & 0.6810 & 0.5320 & 0.5522 & {\scriptsize\textcolor{darkgreen}{$\mathbf{+3.5\%}$}}
& 0.6108 & 0.6742 & 0.5261 & 0.5467 & {\scriptsize\textcolor{darkgreen}{$\mathbf{+2.4\%}$}} \\
\cline{2-13}

& \multirow{2}{*}{Others}
& LONGER$^{\dagger}$
& 0.5607 & 0.6354 & 0.4688 & 0.4930 & {\scriptsize\textcolor{red}{$-2.5\%$}}
& -- & -- & -- & -- & -- \\
& & PersRec
& \textbf{0.6266} & \textbf{0.6920} & 0.5325 & \textbf{0.5543} & {\scriptsize\textcolor{darkgreen}{$\mathbf{+3.7\%}$}}
& \textbf{0.6477} & \textbf{0.7034} & 0.5510 & 0.5703 & {\scriptsize\textcolor{darkgreen}{$+4.7\%$}} \\
\hline

\multirow{8}{*}{XLong}
& \multirow{3}{*}{Original}
& Full
& 0.3024 & 0.4148 & 0.2350 & 0.2709 & {\scriptsize\textit{ref}}
& 0.4057 & 0.4809 & 0.3198 & 0.3441 & {\scriptsize\textit{ref}} \\
& & Short
& 0.2701 & 0.3744 & 0.2010 & 0.2380 & {\scriptsize\textcolor{red}{$-3.3\%$}}
& 0.3255 & 0.4343 & 0.2585 & 0.2932 & {\scriptsize\textcolor{red}{$-5.1\%$}} \\
& & +prefix
& 0.2826 & 0.3911 & 0.2121 & 0.2467 & {\scriptsize\textcolor{red}{$-2.4\%$}}
& 0.3378 & 0.4494 & 0.2702 & 0.3058 & {\scriptsize\textcolor{red}{$-3.8\%$}} \\
\cline{2-13}

& \multirow{3}{*}{Ours}
& Full
& \textbf{0.4296} & \textbf{0.5038} & \textbf{0.3412} & \textbf{0.3656} & {\scriptsize\textcolor{darkgreen}{$\mathbf{+9.5\%}$}}
& \textbf{0.4224} & \textbf{0.4988} & \textbf{0.3327} & \textbf{0.3581} & {\scriptsize\textcolor{darkgreen}{$+1.4\%$}} \\
& & Short
& 0.3947 & 0.4679 & 0.3155 & 0.3391 & {\scriptsize\textcolor{darkgreen}{$\mathbf{+6.8\%}$}}
& 0.3934 & 0.4710 & 0.3073 & 0.3324 & {\scriptsize\textcolor{red}{$-1.2\%$}} \\
& & Adapted
& 0.4206 & 0.5034 & 0.3271 & 0.3539 & {\scriptsize\textcolor{darkgreen}{$\mathbf{+8.3\%}$}}
& 0.4111 & 0.4911 & 0.3202 & 0.3461 & {\scriptsize\textcolor{darkgreen}{$\mathbf{+0.2\%}$}} \\
\cline{2-13}

& \multirow{2}{*}{Others}
& LONGER$^{\dagger}$
& 0.3025 & 0.3937 & 0.2206 & 0.2500 & {\scriptsize\textcolor{red}{$-2.1\%$}}
& -- & -- & -- & -- & -- \\
& & PersRec
& 0.3896 & 0.4761 & 0.3008 & 0.3287 & {\scriptsize\textcolor{darkgreen}{$+5.8\%$}}
& 0.3604 & 0.4776 & 0.2918 & 0.3292 & {\scriptsize\textcolor{red}{$-1.5\%$}} \\
\hline

\end{tabular}%
}
\end{table*}

Table~\ref{tab:main-results} reports the evaluation results across three datasets and two backbone models. Notably, the $\Delta$ column measures the absolute change in NDCG@10 relative to the \textit{Full} setting of each model's original training recipe (serving as the baseline for each dataset). Consequently, a positive value indicates a performance improvement, whereas a negative value represents a decrease relative to this baseline configuration.

\textbf{Performance degradation and the prefix effect under truncation.} 
Within the \textit{Original} group, the \textit{Full} setting serves as the baseline ($ref$). When restricting the inference window to the 20 most recent items (\textit{Short}), performance drops significantly across all configurations, while the \textit{+prefix} setting consistently mitigates this drop. This trend perfectly aligns with our empirical analysis and observations regarding historical visibility in Sec.~\ref{zs_infer}.

\textbf{Advantages of the long-view training recipe.}
Our proposed long-view backbone training recipe (\textit{Ours Full}) achieves a substantial performance elevation, consistently outperforming \textit{Original Full} across all dataset-backbone pairs. This distinct gap demonstrates that our recipe effectively pushes the performance ceiling of the backbone architectures, validating the design philosophy detailed in Sec.~\ref{sec:amplified_sparse}.
\begin{table*}[t]
\centering
\caption{Effectiveness and system overhead on Taobao (HSTU backbone). \textbf{Bold} marks the better entry in each column.}
\label{tab:hstu_compare}
\small
\setlength{\tabcolsep}{5pt}
\renewcommand{\arraystretch}{1.25}
\resizebox{\textwidth}{!}{%
\begin{tabular}{@{}l cc | cccc @{}}
\toprule
\multirow{2}{*}{\textbf{Method}}
  & \multicolumn{2}{c|}{\textbf{Effectiveness (Taobao)}}
  & \multicolumn{4}{c}{\textbf{System Overhead}} \\
\cmidrule(lr){2-3}\cmidrule(l){4-7}
& NDCG@10 & Hit@10
& Trainable Params & Training Time & Long-term Memory & Prefill Cost \\
\midrule

HSTU minimal adaptation (ours)
& $0.5467$ & $0.6742$
& $\mathbf{2{,}764}$
& $\mathbf{3{,}842\text{s}\ (1.07\text{h})}$
& $\mathbf{none}$
& $\mathbf{none}$ \\

HSTU PersRec full-ft
& $\mathbf{0.5703}$ & $\mathbf{0.7034}$
& $302{,}825{,}997\ {\scriptstyle(\times 109{,}600\uparrow)}$
& $30{,}021\text{s}\ (8.34\text{h})\ {\scriptstyle(\times 7.8\uparrow)}$
& $16\ \text{KB/user}\ (1.44\ \text{GiB total})$
& $190\ \text{tokens/user}$ \\

\bottomrule
\end{tabular}%
}
\end{table*}

\textbf{Strong short-view adaptability via zero-shot inference.} 
Remarkably, without any secondary adaptation, \textit{Ours Short} (which directly performs inference on short sequences using the long-view trained weights) already surpasses or matches the performance of \textit{Original Full} in nearly all scenarios. This zero-shot transfer capability indicates that our long-view recipe implicitly cultivates robust short-view adaptability within the learned representation space.

\textbf{Efficacy of short-view adaptation and unexpected gains.} 
The \textit{Adapted} setting, which introduces a fine-tuning stage on short sequences, consistently delivers further performance improvements, largely reclaiming the capacity of \textit{Ours Full}. Notably, on the Taobao dataset under the SASRec backbone, \textit{Adapted} even outperforms \textit{Ours Full} in terms of Hit@5 ($0.6190$ vs. $0.6144$) and Hit@10 ($0.6810$ vs. $0.6677$), suggesting that short-view fine-tuning can recover and in some cases exceed the full-view performance ceiling.

\textbf{Superiority over alternative caching baselines.} 
Compared to external baselines, our approach demonstrates a robust and unified advantage. It consistently outperforms LONGER$^\dagger$, which is trained from scratch without the long-view backbone warm-start that all other methods in this comparison share. When compared with PersRec, although PersRec exhibits competitive performance on Taobao, it inherently demands a full-parameter fine-tuning protocol. In contrast, our method achieves superior or highly competitive results while updating only a minimal fraction of parameters. Furthermore, on the ML-10M and XLong datasets, our method consistently dominates the warm-started PersRec variant. A detailed efficiency and parameter overhead analysis is provided in Sec.~\ref{sec:efficiency}.

\subsection{Effectiveness--Efficiency Trade-off}\label{sec:efficiency}

Sec.~\ref{exp_overall} shows that our method is competitive with PersRec in recommendation quality, but the two approaches differ fundamentally in what they require at training and serving time. PersRec fine-tunes all parameters and maintains a per-user token cache; our method updates only bias and LayerNorm parameters and needs no auxiliary storage. Table~\ref{tab:hstu_compare} makes this trade-off concrete by comparing our HSTU minimal-adaptation model against HSTU PersRec full-ft on Taobao across both effectiveness and system overhead.

As Table~\ref{tab:hstu_compare} shows, PersRec's information advantage---a pre-filled KV cache of $190$ historical tokens per user---yields higher NDCG@10 ($0.5703$ vs.\ $0.5467$). The efficiency picture reverses sharply: our method updates only $2{,}764$ parameters ($109{,}000\times$ fewer than PersRec's $302$M), trains in $1.07$h versus $8.34$h, and imposes no long-term memory or per-user prefill at serving time. In total, our method reaches $95.9\%$ of PersRec's NDCG@10 while eliminating all per-user state infrastructure.

\subsection{Ablation Studies}
We conduct ablations aligned with the three design choices in Sec.~\ref{sec:method}: the long-view backbone recipe, the architecture-aware minimal-adaptation rule, and the dual-supervision short-view objective.
\subsubsection{Long-View Backbone Ablation}
\begin{table}[t]
\centering
\small
\setlength{\tabcolsep}{5pt}
\renewcommand{\arraystretch}{1.1}
\caption{Step-wise backbone recipe ablation on Taobao.}
\label{tab:ablation_recipe}
\resizebox{\linewidth}{!}{%
\begin{tabular}{llcc}
\toprule
Model & Recipe Change & NDCG@10 & Hit@10 \\
\midrule
\multirow{3}{*}{SASRec}
& Original (dot + BCE) & 0.5175 & 0.6479 \\
& + similarity + sampled softmax & 0.5435 & 0.6613 \\
& Final (+ $\operatorname{softmax}_{1}$) & \textbf{0.5500} & \textbf{0.6677} \\
\midrule
\multirow{4}{*}{HSTU}
& Base recipe (softmax + sampled softmax) & 0.5229 & 0.6652 \\
& + $\operatorname{softmax}_{1}$ & 0.5361 & 0.6718 \\
& + \texttt{rel\_bias} & 0.5615 & 0.6814 \\
& Final (+both) & \textbf{0.5764} & \textbf{0.6992} \\
\bottomrule
\end{tabular}%
}
\end{table}

Table~\ref{tab:ablation_recipe} validates the design choices in our long-view backbone training recipe. The results show that the performance gain accumulates through a structured recipe progression rather than a single modification, though the dominant refinements differ across architectures.

For SASRec, transitioning from raw dot-product scoring with BCE loss to cosine similarity under sampled softmax yields the most substantial improvement. Because similarity scoring and sampled softmax are formulated to complement each other to maximize representation alignment, we evaluate them as a combined configuration rather than in isolation. Additionally, incorporating \texttt{softmax1} on top of this combination delivers a further boost, confirming that the synthesis of both techniques yields the most notable performance enhancement for SASRec.

HSTU follows a distinct trajectory, which we evaluate across its two primary official implementation pathways. The first path utilizes standard Softmax attention: introducing \texttt{softmax1} into this baseline yields a moderate but consistent performance gain. The second, recommended path replaces standard attention with relative position bias (\texttt{rel-bias}) and adopts the SiLU activation~\cite{elfwing2018sigmoid}. Under this stronger baseline pathway, further integrating \texttt{softmax1} produces the most pronounced enhancement, achieving the peak performance of 0.5762. These results demonstrate that while the critical component varies by backbone, each architecture consistently benefits from the geometry-aware recipe.
\subsubsection{Short-View Parameter Ablation}
\begin{table}[t]
\centering
\small
\setlength{\tabcolsep}{5pt}
\renewcommand{\arraystretch}{1.1}
\caption{Architecture-aware parameter-family ablation for short-view adaptation on Taobao.}
\label{tab:ablation_finetune}
\begin{tabular*}{\linewidth}{@{\extracolsep{\fill}}llcc@{}}
\toprule
Model & Trainable Params & NDCG@10 & Hit@10 \\
\midrule
\multirow{4}{*}{SASRec}
& Direct short & 0.5406 & 0.6511 \\
& + bias only & 0.5492 & 0.6770 \\
& + LayerNorm only & 0.5501 & 0.6767 \\
& + Both & \textbf{0.5520} & \textbf{0.6805} \\
\midrule
\multirow{4}{*}{HSTU}
& Direct short & 0.5396 & 0.6631 \\
& + bias only & \textbf{0.5472} & \textbf{0.6748} \\
& + LayerNorm only & 0.5375 & 0.6623 \\
& + Both & 0.5458 & 0.6733 \\
\bottomrule
\end{tabular*}
\end{table}

Table~\ref{tab:ablation_finetune} evaluates the fine-tuning efficiency of different parameter families. The \textit{Direct short} baseline applies the long-view trained backbone to short sequences without adaptation. 

The results show that fine-tuning either biases or LayerNorm (LN) parameters improves performance over the baseline, but the impacts are architecture-dependent. For SASRec, LN-only adaptation is more effective than bias-only because SASRec relies heavily on parameterized LN layers. Combining both parameter families yields the best performance ($0.5520$). 

Conversely, HSTU favors bias-only adaptation, which captures almost all the gain ($0.5472$), while LN-only adaptation contributes little. This aligns with HSTU's architecture, where block normalizations are parameter-free, and trainable parameters concentrate within its relative-position and projection biases. Overall, jointly fine-tuning both parameter families remains a robust and stable choice for both backbones.

\subsubsection{Loss Ablation}

\begin{table}[t]
\centering
\small
\setlength{\tabcolsep}{5pt}
\renewcommand{\arraystretch}{1.1}
\caption{Short-view supervision objective ablation on two representative regimes.}
\label{tab:loss_ablation}
\begin{tabular*}{\linewidth}{@{\extracolsep{\fill}}lcc|cc@{}}
\toprule
\multirow{2}{*}{Loss} & \multicolumn{2}{c|}{HSTU / Taobao} & \multicolumn{2}{c}{SASRec / XLong} \\
& NDCG@10 & Hit@10 & NDCG@10 & Hit@10 \\
\midrule
Direct short & 0.5396 & 0.6631 & 0.3391 & 0.4679 \\
BCE only & \textbf{0.5467} & \textbf{0.6742} & 0.3539 & 0.5034 \\
KD only & 0.5462 & 0.6722 & \textbf{0.3682} & \textbf{0.5169} \\
BCE + KD & 0.5461 & 0.6738 & 0.3543 & 0.5040 \\
\bottomrule
\end{tabular*}
\end{table}

Table~\ref{tab:loss_ablation} evaluates the dual-supervision objective $\mathcal{L}_{\mathrm{FT}}$ from Eq.~\eqref{eq:method-ft-final}, with the bias+LayerNorm parameter set fixed across all variants. On Taobao with HSTU, BCE-only supervision achieves the strongest result, and adding the candidate-set distillation term $\mathcal{L}_{\mathrm{KD}}$ brings no further improvement. On XLong with SASRec, the picture reverses: $\mathcal{L}_{\mathrm{KD}}$ alone outperforms BCE-only ($0.3682$ vs.\ $0.3539$ NDCG@10), indicating that the teacher's relative candidate ordering provides a valuable signal when the long-to-short truncation gap is larger.

These results confirm that the optimal supervision balance is regime-dependent. The scalar $\lambda_{\mathrm{KD}}$ in Eq.~\eqref{eq:method-ft-final} acts as a regime-sensitive control: direct BCE anchoring suffices where the short-view performance gap is moderate, while candidate-set distillation from the full-history teacher provides complementary structural signal in more challenging truncation settings.

\subsection{Temporal Robustness Under Rolling Evaluation}

\paragraph{Protocol.}
Motivated by the decay analysis in PersRec~\cite{zhang2025efficient}, we design a rolling evaluation to test whether adaptation gains persist across a stream of future requests. Specifically, we hold out the last ten (instead of last one in LOO settings)
interactions as a future block (reducing the effective training length
to 190 items on Taobao and 390 on XLong). The target at position $-10$ is reserved for model selection, and we report NDCG@10 at positions $-9,\dots,-1$. At each position, the model observes only the sliding window of the 20 most recent interactions before that request.

\begin{figure}[t]
    \centering
    \includegraphics[width=\columnwidth]{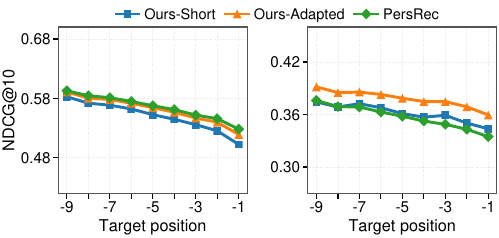}
    \caption{Rolling NDCG@10 over nine future targets. \textbf{Ours-Short} uses direct short-view inference, \textbf{Ours-Adapted} adds short-view adaptation, and \textbf{PersRec} uses a prefilled token cache. Each method observes the latest 20 interactions: (a)~SASRec on Taobao; (b)~HSTU on XLong.}
    \Description{Two line charts compare Ours-Short, Ours-Adapted, and PersRec across nine future requests. On Taobao with SASRec, all methods decline similarly and PersRec remains slightly ahead. On XLong with HSTU, Ours-Short leads on average and Ours-Adapted consistently exceeds PersRec.}
    \label{fig:rolling-main}
\end{figure}

\paragraph{Compared methods.}
Figure~\ref{fig:rolling-main} compares \textbf{Ours-Short} (the long-view backbone under direct short-view inference), \textbf{Ours-Adapted} (the same backbone after short-view fine-tuning), and \textbf{PersRec} (eight personalized tokens prefilled from 180/382 historical items on Taobao/XLong). Both our methods serve using only the 20 most recent items.

\paragraph{Results.}

On Taobao with SASRec (Figure~\ref{fig:rolling-main}(a)), all three curves follow a similar declining trend from position $-9$ to $-1$, and their relative ordering is stable throughout. PersRec leads, Ours-Adapted is consistently second, and Ours-Short is weakest. Rolling averages over the nine positions are $0.5494$/$0.6622$, $0.5607$/$0.6848$, and $0.5652$/$0.7065$ in NDCG@10/Hit@10, respectively. Despite operating without the long-term cache that PersRec relies on, Ours-Adapted recovers $97.4\%$ of PersRec's average NDCG@10, and the gap remains stable rather than widening as the target moves further into the future block.

On XLong with HSTU (Figure~\ref{fig:rolling-main}(b)), the picture shifts. Ours-Adapted achieves the highest rolling average ($0.3780$/$0.5232$) and leads at all nine target positions, demonstrating that the benefit of short-view adaptation persists throughout the rolling horizon. Ours-Short reaches $0.3616$/$0.5099$ on average, slightly exceeding PersRec's $0.3571$/$0.5051$ despite the latter's prefilled cache.

\paragraph{Takeaway.}
Across both datasets and all rolling positions, Ours-Adapted matches or outperforms PersRec while using only the recent context at serving time. This demonstrates that our short-view adaptation is not narrowly tuned to a single terminal target but delivers temporally stable gains throughout an evolving request stream—without the per-user memory infrastructure that PersRec requires.

\subsection{Embedding Visualization}

\begin{figure}[t]
    \centering
    \includegraphics[width=1\linewidth]{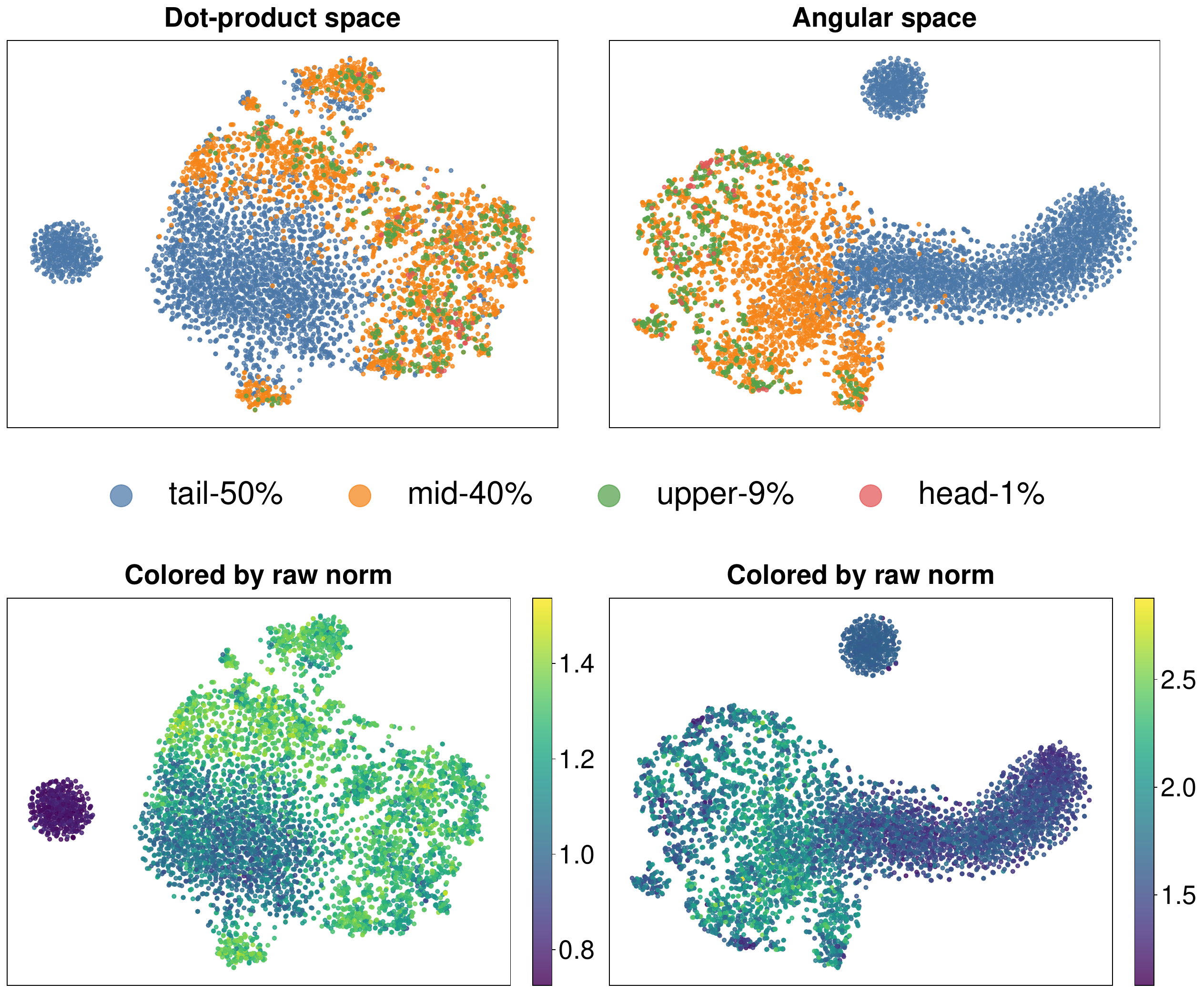}
    \caption{t-SNE of Taobao item embeddings (SASRec). Left: dot-product; Right: similarity-based. Top: popularity tier; Bottom: raw embedding norm.}
    \Description{A two-by-two t-SNE comparison of Taobao item embeddings. The left column uses dot-product scoring and shows heavily intermixed popularity tiers and irregular norms. The right column uses similarity-based scoring and shows clearer popularity structure together with a smoother norm gradient.}
    \label{fig:tsne}
\end{figure}

Figure~\ref{fig:tsne} visualizes Taobao item embeddings by popularity tier and raw norm to examine how angular scoring changes the norm--popularity structure identified in Sec.~\ref{sec:amplified_sparse}.

\paragraph{Popularity structure.}
Dot-product embeddings heavily intermix the four popularity tiers, whereas angular scoring produces a clearer frequency-ordered structure. Both spaces retain an isolated long-tail cluster, suggesting that these niche items remain distinct under either scoring regime. In the angular space, head and upper-frequency items occupy one end of the main structure, mid-frequency items lie centrally, and tail items extend along its wider end. This separation is consistent with reduced reliance on global popularity in the learned representation.

\paragraph{Norm structure.}
Under dot-product scoring, norms are irregularly distributed within a narrow range ($0.8$--$1.5$), including isolated high-norm regions that can act as popularity-correlated shortcuts through Eq.~\eqref{eq:analysis-dot-decomp}. Angular scoring yields a smoother gradient from head/upper-frequency regions ($\ge 2.5$) to the mid- and tail-frequency regions, including the low-norm tail island ($\approx 1.2$). Because magnitude is removed from ranking, these norm differences no longer directly determine candidate order.

\section{Related Work}

\subsection{Long-Context Recommendation}

Sequential recommendation predicts the next item from a user's historical interactions. Early neural methods use recurrent architectures such as GRU4Rec~\cite{hidasi2015session}, while later methods adopt self-attention or bidirectional sequence modeling, including SASRec~\cite{kang2018self} and BERT4Rec~\cite{sun2019bert4rec}. Recent efficient backbones further improve scalability through linear attention, recurrent units, state-space models, or transducer-style architectures, such as LRURec~\cite{yue2024linear}, Mamba4Rec~\cite{liu2024mamba4rec}, and HSTU~\cite{zhai2024actions}. Beyond standard sequential recommendation, long-context recommendation aims to exploit extended user histories for richer interest modeling. Industrial studies usually handle ultra-long histories by sampling, clustering, compressing, or retrieving informative behaviors, as in TWIN~\cite{chang2023twin}, TWIN V2~\cite{si2024twin}, and LONGER~\cite{chai2025longer}. Other studies improve long-sequence efficiency through scalable attention or sequence compression mechanisms~\cite{feng2024long,hsieh2024ruler}. These methods demonstrate the value of long histories, but they generally assume that long-range information, or a compressed version of it, is available during inference. In contrast, our work focuses on cache-cold serving, where the model is trained with long histories but must serve from only a short recent window.

\subsection{Efficient Serving and Adaptation}

A related direction compresses long contexts into compact states to reduce inference cost. In language models, prompt and context compression methods introduce learnable tokens or autoencoding objectives to summarize long inputs, such as Gist Tokens~\cite{mu2023learning}, SepLLM~\cite{chen2024sepllm}, In-context Autoencoder~\cite{ge2023context}, and recent analyses of gist-token compression~\cite{deng2025silver}. Similar ideas have been explored in recommendation: PersRec~\cite{zhang2025efficient} compresses long-term user histories into personalized states, Kwai Summary Attention~\cite{chu2026kwai} summarizes user behaviors into compact attention states, Recurrent Preference Memory~\cite{chen2026recurrent} maintains recurrent user memories, and MTServe~\cite{wang2026mtserve} studies hierarchical caches for generative recommendation serving. Although these methods reduce repeated long-history computation, they still require history-preserving states to be precomputed, stored, refreshed, or retrieved during serving. Our method is complementary: instead of maintaining per-user memory or KV caches, we adapt the backbone itself to operate under recent-window inference.

Our work is also related to studies on retrieval geometry and training objectives. Most ID-based recommenders use inner-product scoring, following matrix factorization and neural collaborative filtering~\cite{rendle2020neural,he2017neural}. However, dot-product scoring entangles angular similarity with vector magnitude, and prior studies show that embedding norms may encode popularity or frequency shortcuts~\cite{zhai2023revisiting}. Recent analyses further suggest that sequential recommendation performance can be strongly affected by loss functions and evaluation protocols~\cite{klenitskiy2023turning,petrov2023gsasrec}. Motivated by these findings, we train the long-view backbone with normalized angular scoring and sampled-softmax competition, then perform lightweight short-view adaptation by updating only native bias and LayerNorm parameters.
\section{Conclusion and Future Work}

In this paper, we studied the long-short-view performance gap in sequential recommendation, motivated by a practical deployment constraint: serving systems often restrict inference to a short recent window while distant history requires prefilling, per-user storage, or cache retrieval that may not always be available. We argue that this gap can be addressed largely on the training side, without reconstructing the full history at every request.

Through empirical analysis, we identify two structural flaws—dot-product norm bias and a learned dependency on prefix positions that breaks upon truncation—and show that correcting them via angular scoring and softmax1 already surpasses conventional full-view baselines under zero-shot short-view inference. Lightweight bias/LayerNorm adaptation with cross-view distillation then closes the remaining gap at minimal parameter cost, eliminating all per-user memory and prefill overhead. Both stages operate on components that standard backbones already expose, making the framework applicable across architectures.

We plan to extend this work in three directions: scaling to even longer sequence lengths; exploring how semantic and multimodal signals can help bridge the long-short-view gap; and investigating it within LLM-augmented and generative recommender systems.

\section*{GenAI Usage Disclosure}
OpenAI Codex with GPT-5.4 and GPT-5.5 was used to assist with code refactoring, debugging, documentation, and verification of the experimental results. Claude Code with Claude Opus 5.6 was used to assist with manuscript organization and figure refinement. All AI-assisted outputs were reviewed and verified by the authors, who take responsibility for the paper's content, figures, analyses, and conclusions.

\bibliographystyle{ACM-Reference-Format}
\bibliography{sample-base}

\end{document}